\documentclass{pasj00}
\draft

\begin{document}
\SetRunningHead{Kawanaka et al.}{X-ray emissions from MHD accretion flows}

\title{X-ray Emissions from Three-dimensional Magnetohydrodynamic Coronal
 Accretion Flows}
\author{Norita \textsc{Kawanaka},\altaffilmark{1}
 Yoshiaki \textsc{Kato},\altaffilmark{2}
 and Shin \textsc{Mineshige}\altaffilmark{1}}
\altaffiltext{1}{Yukawa Institute for Theoretical Physics, Kyoto University,
Kyoto 606-8502, Japan}
\altaffiltext{2}{Center for Computational Sciences,
 University of Tsukuba 1-1-1 Tennodai, Tsukuba, Ibaraki 305-8511, Japan}
\email{norita@yukawa.kyoto-u.ac.jp}
\KeyWords{accretion, accretion disks --- black hole physics ---
 radiative transfer -- X-rays: general}

\maketitle
\begin{abstract}
We calculate the radiation spectrum and its time variability of the black
 hole accretion disk-corona system
 based on the three-dimensional magnetohydrodynamic simulation. 
 In explaining the spectral properties of active galactic nuclei (AGNs),
 it is often assumed
 that they consist of a geometrically thin, optically thick disk and hot,
 optically thin corona surrounding the thin disk.
 As for a model of the corona, we adopt the simulation data of
 three-dimensional, non-radiative MHD accretion flows 
 calculated by Kato and coworkers, while for a thin disk we assume
 a standard type disk.  We perform
 Monte Carlo radiative transfer simulations in the corona, taking into
 account the Compton scattering of soft photons 
 from the thin disk by hot thermal electrons and coronal irradiation heating
 of the thin disk, which emits blackbody radiation.
  By adjusting the density parameter of the MHD coronal flow, 
 we can produce the emergent spectra which are consistent with those of
 typical Seyfert galaxies.  Moreover, we find 
 rapid time variability in X-ray emission spectra, originating from the
 density fluctuation produced by the magnetorotational instability
 in the MHD corona.  
 The features of reflection component including iron fluorescent line
 emission are also briefly discussed.
\end{abstract}

\section{Introduction}
Thanks to the rapid progress in the observational studies in recent years,
 our understanding of the radiation properties of black hole accretion flows
 have been
 really deepened.  Accreting black holes, such as active galactic nuclei
 (AGNs) and black hole binaries (BHBs) during their very-high
 spectral state [state with luminosities around a few
 tenths of the Eddington limit ($L_{\rm Edd}$)], show the radiation spectra
 dominated by two components; the thermal bump in UV/soft X-ray
 band and the power-law emission with a spectral index of $\alpha\sim 1$
 in the X-ray band (possibly with a high energy cutoff around MeV; for
 the multiwavelength spectrum of a typical AGN, see Reynolds et al. 1997).  
 These components are often explained by the disk-corona model
 (Liang \& Price 1975; Bisnovatyi-Kogan \& Blinnikov 1977;
 Haardt \& Maraschi 1991, 1993).  In this model, the accretion flow consists
 of geometrically thin, optically thick accretion disk whose structure is
 studied by Shakura \& Sunyaev (1973), and hot, optically thin corona
 surrounding the disk.  The thermal bump is believed as the thermal emission
 from the optically thick disk (see Kishimoto et al. 2005 for its
 observational implication), and the power-law component is interpreted
 to be formed by photons which are emitted from
 the disk and Compton up-scattered by hot electrons in the corona
 (see Thorne \& Price 1975; Shapiro et al. 1976 but in the context of the
 inner hot accretion flow model).  Such a structure that the hot gas
 (i.e. the corona) coexists with the cool gas
 (i.e. the disk) is justified by the existence of the reflection component
 observed in the X-ray spectra  (Pounds et al. 1990; see also
 Guilbert \& Rees 1988; Lightman and White 1988).  
 This reflection component is accompanied by the iron fluorescent line
 emission broadened relativistically (for reviews, see Mushotzky et al. 1993;
 Fabian et al. 2000; Reynolds \& Nowak 2003).

On the other hand, our theoretical understanding of black hole accretion has
 also made a rapid progress.  Now the dynamics of accretion flows is
 understood in terms of magnetohydrodynamics (MHD), since Balbus \& Hawley
 (1991) rediscovered the magnetorotational instability (MRI) as the
 fundamental mechanism of angular momentum transfer in accretion disks.
  Detailed dynamical features of MHD accretion flows have been investigated
 via global three-dimensional numerical simulations (Matsumoto 1999;
 Stone \& Pringle 2001; Machida et al. 2001; Hawley \& Krolik 2001, 2002;
 Machida \& Matsumoto 2003; Armitage \& Reynolds 2003;
 De Villiers et al. 2003; Gammie et al. 2003; Igumenshchev et al. 2003).
 Such magnetically dominated accretion flows as those simulated in these
 calculations are radiatively inefficient accretion flows (RIAFs) 
 whose mass accretion rates are much smaller than the critical value of
 $\sim L_{\rm Edd}$ (see, e.g., Mineshige et al. 1995).  
 In such flows, the dissipated energy would not be radiated away
 efficiently, because of their low density, and be advected inward to the
 central black hole (Ichimaru 1977; Narayan \& Yi 1994;
 Abramowicz et al. 1995; Kato et al. 1998).  

Although the detailed structures and behavior of magnetohydrodynamical
 accretion flows are shown by numerical methods, it is still unclear
 that if these simulational results can be applied in realistic
 situations in the universe.  In recent years, the researches concerning
 the observational properties like the spectra and their time
 variabilities expected from numerically simulated accretion flows have
 been extensively performed.  Since the RIAF model is believed to fit the
 emergent spectrum of Sagittarius ${\rm A}^*$, some authors have
 calculated the time dependent radiation spectra predicted from simulated
 MHD accretion flows with the aim to reproduce the observed behavior of
 Sgr ${\rm A}^*$ (Hawley \& Balbus 2002; Goldstone et al. 2005;
 Ohsuga et al. 2005; Moscibrodzka et al. 2007).  However, as for the
 accretion flows with moderately high mass accretion rate, which are
 considered to be a good model for AGNs and BHBs showing the thermal bump
 and the power-law emission in their spectra, no attempts have been made
 so far to calculate the spectra predicted from MHD simulations taking
 into account realistic radiation processes.  As noted in the above, in
 order to reproduce such spectra, the simulated accretion flows should
 have two components: geometrically thin cool disk and optically thin
 hot corona.  However, no simulation data which incorporate such
 two-component effects are available.

In this study we calculate for the first time the emergent spectra of
 two-component accretion flows based on the three-dimensional MHD
 simulation by Kato et al. (2004, hereafter KMS04).  Instead of fully
 solving the dynamics of two-component accretion flows, we adopt the
 simulation result of RIAF-like
 MHD accretion flow for the optically thin corona, and assume that
 the optically thick, geometrically thin disk is embedded in the corona
 with mutual interaction through radiation.  Other interactions,
 such as mass evaporation/condensation or heat conduction, are
 neglected, for simplicity (Meyer \& Meyer-Hofmeister 1994; Meyer
 et al. 2000; Liu et al. 2002, 2007).  The disk is emitting soft
 photons with thermal spectrum, and those photons are up-scattered
 by hot electrons in the corona.  After being scattered in the corona,
 a part of upscattered photons return to the disk and are
 thermalized there, thereby heating the accretion disk.  The disk
 emission is, hence, enhanced by this returning process.
 In this paper we perform three-dimensional Monte Carlo radiative
 transfer simulations to properly calculate  such radiation processes
 and the emergent spectra.

The plan of this paper is as follows.  We show the detail of the model
 and method used in our calculation in \S 2.  The results are presented in
 \S 3, and compare them with the spectra observed in typical AGNs
 and discuss the similar points and discrepancies between them in \S 4.
  In \S 5 we summarize our study.

\section{MODEL AND CALCULATION METHODS}
\subsection{Overview of Adopted MHD Simulations}
KMS04 investigated the evolution of a torus threaded by weak localized
 poloidal magnetic fields by performing the three-dimensional MHD
 simulation.  They solved the following basic equations of the resistive
 MHD in the cylindrical coordinates, $(r,~\phi ,~z)$:
\begin{eqnarray}
\frac{\partial \rho}{\partial t}+\mbox{\boldmath $\nabla$}\cdot
 (\rho \mbox{\boldmath $v$})=0, \\
\frac{\partial}{\partial t}(\rho \mbox{\boldmath $v$})+
\mbox{\boldmath $\nabla$}\cdot \left(\rho \mbox{\boldmath $v$}\otimes
 \mbox{\boldmath $v$}
 -\frac{\mbox{\boldmath $B$}\otimes \mbox{\boldmath $B$}}{4\pi} \right)
 = -\mbox{\boldmath $\nabla$}\left(p_{\rm gas}+\frac{B^2}{8\pi}\right)
-\rho\mbox{\boldmath $\nabla$}\psi ,\\
\frac{\partial}{\partial t}\left(\varepsilon +\frac{B^2}{8\pi}\right) 
+\mbox{\boldmath $\nabla$}\cdot \left[\left(\varepsilon 
+p_{\rm gas}\right)\mbox{\boldmath $v$}
 +\frac{c(\mbox{\boldmath $E$}\times \mbox{\boldmath $B$})}{4\pi}\right]
=-\rho\mbox{\boldmath $v$}\cdot \mbox{\boldmath $\nabla$}\psi, \\
\frac{\partial B}{\partial t}=-c\mbox{\boldmath $\nabla$}\times
 \mbox{\boldmath $E$},
\end{eqnarray}
where $\psi=-GM/(R-r_{\rm S})$ is the pseudo-Newtonian potential
 (Paczy\'{n}ski \& Wiita 1980), $\varepsilon =\rho v^2/2+p_{\rm gas}/
(\gamma-1)$ is the energy of the gas (here $\gamma$ is fixed to be 5/3)
 and $\mbox{\boldmath $E$}=
-(\mbox{\boldmath $v$}/c)\times \mbox{\boldmath $B$}+(4\pi\eta/c^2)
\mbox{\boldmath $J$}$ is Ohm's law.  Here $R~[\equiv (r^2+z^2)^{1/2}$
 is the distance
 from the origin,$r_{\rm S}~(\equiv 2GM/c^2)$ is the Schwarzschild
 radius (with $M$ and $c$ being the mass of a BH and the speed of
 light, respectively),
 and $\mbox{\boldmath $J$}=(c/4\pi)
\mbox{\boldmath $\nabla$}\times \mbox{\boldmath $B$}$ is the electric
 current.  As to $\eta$ we adopt the anomalous resistivity model which
 is used in many
 solar flare simulations (for the detail, see Yokoyama \& Shibata 1994).
  The calculation is started with a rotating torus in hydrostatic balance
 located around $r=r_0=40r_{\rm S}$.  The initial magnetic fields are
 confined within a torus and purely poloidal (see KMS04 for the detail).
  In the simulation
 they used $300\times 32\times 400$ nonuniform mesh points.  The grid
 spacing is uniform $(\Delta r=\Delta z=0.16r_{\rm S})$ within the inner
 calculation box of
 $0\leq z\leq 10r_{\rm S}$, and it increases by 1.5\% from one mesh to
 the adjacent outer mesh outside this box up to $r\leq 20r_{\rm S}$ and
 $z\leq 20r_{\rm S}$, and it
 increases by 3\% beyond that.  The entire computational box size is
 $0\leq r\leq 200r_{\rm S}$, $0\leq \phi \leq 2\pi$,
 and $-50r_{\rm S}\leq z\leq 50r_{\rm S}$, and they simulated a full
 $360^\circ $ domain (see Kato 2004 for more detail).

 The simulated MHD flow is slightly oscillating because of the turbulence
 driven by MRI, and geometrically thick density distribution is produced.
  In this quasi-steady accretion flow,
 the density profile is $\rho \propto r$ in the inner part
 ($r<20r_{\rm S}$), while $\rho \propto r^{-1}$ in the outer
 part ($r>20r_{\rm S}$) (see Fig. 4 in KMS04).  We use this quasi-steady
 density distribution as well as the ion temperature distribution in
 modeling the corona in which hot thermal electrons up-scatter the soft
 photons emerging from the cold disk virtually located in the equatorial
 plane.

\subsection{Physical Quantities of the Disk and Corona} 
As mentioned in \S 1 briefly, we adopt the following assumptions in
 constructing the disk-corona model.

1. In the equatorial plane, we assume a standard accretion disk
 (Shakura \& Sunyaev 1973) with an infinitesimal height for a given
 mass accretion rate $\dot{M}_{\rm disk}$.

2. The RIAF-like accretion flow, whose structure is provided by the
 three-dimensional MHD simulation (KMS04), is surrounding the standard
 thin disk
 as the hot corona.  That is, we assume that coronal flow is created at
 large radius by evaporation of the disk material and moves freely
 inward
\footnote{Note that evaporation dominates over condensation at large
 radii, whereas the opposite is the case at small radii
 (e.g. Liu et al. 2007)}.
  Mass evaporation and condensation occurring between them and heat
 conduction from the corona to the disk are neglected in the calculation.
  Radiative
 cooling of coronal plasma is also neglected.

3. The photons radiated from the thin disk are partly up-scattered by
 hot electrons in the corona and the remaining portion penetrate
 through the
 corona, creating a power-law hard emission in the spectrum. 

4. As the seed photon field from the underlying disk, we include the
 reprocessed radiation, which comes from the coronal irradiation
 on the disk, as
 well as the intrinsic disk radiation.

We take three-dimensional data of density and proton temperature
 distributions in the accretion flows calculated in Kato
 (2004). This has the same initial condition with Model B in KMS04.  The data
 of physical properties
 are given at each point in the simulation box associated with Cartesian
 coordinates $(x,~y,~z)$, in which the black hole is located at the
 origin of the
 coordinate axes, the $z$-axis is set to be the rotation axis of the
 accretion flow, and the $x$-$y$ plane corresponds to the equatorial
 plane.  We employ
 Cartesian grids with numbers $(N_x,~N_y,~N_z)=(101,~101,~101)$ of cells.
  The size of the calculating box is $2X\times 2Y\times 2Z$, where we set
$(X,~Y,~Z)=(99.9r_{\rm S},~99.9r_{\rm S},~99.9r_{\rm S})$.

In the MHD simulations with no radiative loss the density is given as
 non-dimensional number ${\tilde \rho}$ with the normalization factor
 $\rho_0$,
 which is treated a free parameter in our calculation.  Basically the
 coronal density is determined by evaporation of the disk gas, but here we
 determine $\rho_0$
 so that the power-law indices of the evaluated spectra agree with
 the observations.  The proton temperature does not depend on the
 density parameter and
 is given by the MHD simulation as
 $(\mu m_p c^2/k_B){\tilde c}_s^2$, where $\mu$ is the mean molecular
 weight $(=0.5)$, $m_p$ is the proton mass, $k_B$ is the Boltzmann constant,
 and ${\tilde c}_s$ is the normalized sound velocity obtained by the
 simulation.

Here we should note that ions (protons) and electrons in the plasma
 simulated in KMS04 have the same temperature, though electrons would be
 radiatively
 cooled and have lower temperature in the realistic situation.  Assuming
 that ion temperature coincides with the simulated one and that the
 electrons
 have a Maxwellian distribution, we evaluate the electron
 temperature, $T_e$, through the energy balance of the electrons between
 Coulomb collisions with ions and radiative cooling,
\begin{eqnarray}
\int_{-Z}^Z \int_{r_{\rm in}}^{r_{\rm out}} \lambda _{ie}2\pi r dr dz
=\int L_C(\nu;~r_{\rm in}\leq r\leq r_{\rm out})d\nu. \label{balance}
\end{eqnarray}
Here $\lambda_{ie}$ is the energy transfer rate from ions to electrons
 (Stepney \& Guilbert 1983) and $L_C(\nu)$ is the coronal luminosity at
 frequency $\nu$.  In the present study, we divide the corona into three
 regions (0<$r\leq 10r_{\rm S}$, $10r_{\rm S}<r\leq 30r_{\rm S}$,
 and $30r_{\rm S}<r$; ($r_{\rm in}$, $r_{\rm out})=(0,10r_{\rm S}),
~(10r_{\rm S},30r_{\rm S})$ and $(30r_{\rm S},\infty )$) for simplicity,
 and suppose that the electrons in the accretion flow
 in each region have the temperature which is independent of the radius
 $r$ and the altitude $z$.  The coronal luminosity, $L_C$, is obtained
 by Monte Carlo simulations
 (see next subsection) for a guess value of $T_e$.  Since this $T_e$
 does not always satisfy Eq. (\ref{balance}), we should do some
 iterations to
 calculate the appropriate electron temperature and the emergent
 spectrum.

\subsection{Radiative Transfer Simulations}
Our model consists of a cold disk which produces blackbody radiation
 at each radius with temperature being determined from the standard
 model (Shakura \& Sunyaev 1973):
\begin{eqnarray}
T^0_{\rm disk}=\left[\frac{3GM\dot{M}_{\rm disk}}{8\pi r^3\sigma}
\left(1-\sqrt{\frac{r}{r_{\rm in}}}\right)\right]^{1/4}, \label{seed}
\end{eqnarray}
 (as long as the reprocessed radiation is unimportant) where $G$
 is the gravitational constant, $\dot{M}_{\rm disk}$ is the mass
 accretion rate in the disk,
 and $\sigma$ is the Stephan-Boltzmann constant.  Note that when
 irradiation flux $F_{\rm irr}$ by the corona is available, the
 disk emits blackbody
 with an enhanced temperature
\begin{eqnarray}
T_{\rm disk}=\left[\left(T_{\rm disk}^0\right)^4+F_{\rm irr}/
\sigma\right]^{1/4}. \label{irr}
\end{eqnarray}
  In the following calculation we set $r_{\rm in}=3r_{\rm S}$,
 $\dot{M}_{\rm disk}=10^{-3}\dot{M}_{\rm Edd}$ (with
 $\dot{M}_{\rm Edd}=10L_{\rm Edd}/c^2$)
 and the mass of a central black hole to be $M=10^8 M_{\odot}$,
 which is believed to be the typical value for AGNs.  Thus, the
 normalized time
 corresponds to ${\tilde t}\cong r_{\rm S}/c=10^3{\rm sec}$.

As the radiation process, we take into account both of the
 intrinsic disk radiation [Eq.(\ref{seed})] and the thermal
 reprocessing from irradiated
 disk as the seed photon field [see Eq. (\ref{irr})], and the
 Compton/inverse Compton scattering in the corona.  We neglect
 synchrotron emission/absorption and free-free emission/absorption.
  Some studies about the
 disk-corona model have shown that the power-law component of the
 emergent spectra is explained as the Comptonized emission
 (Haardt \& Maraschi 1991, 1993)
 , so for our
 present purpose this approximation is justified at least in the
 X-ray energy bands.
  According to the simulation by KMS04, the magnetic field in the
 coronal flow with $\rho_0=1.6\times 10^{-14}~{\rm g}~{\rm cm}^{-3}$
 is $B\sim 10^3{\rm G}$, for which
 the Compton scattering is the most efficient cooling mechanism.
  This fact justifies the method of deriving the coronal temperature
 described
 in the previous section.

The method of the Monte Carlo simulation is based on Pozdnyakov et
 al. (1977).  In order to efficiently calculate the emergent spectra,
 we introduce
 a photon weight $w$.  When emerged
 from the disk we set that each photon has the weight of $w_0=1$,
 and then we calculate
 the escape probability, $P_0$.  The escape probability of a photon
 after
 $i$-th scattering (for $i\geq 1$), $P_i$, is evaluated as
\begin{eqnarray}
P_i=\exp \left( -\int \left[ \frac{\rho(x_i,~y_i,~z_i)}{m_p}\right]
\sigma_{\rm KN}(x_i,~y_i,~z_i) dl\right),
\end{eqnarray}
where $(x,~y,~z)=(x_0,~y_0,~0)$ corresponds to the point on the
 equatorial plane
 (i.e. the disk plane), in which the thermal soft photons are
 generated,
 $(x_i~,y_i,~z_i)$ is the point where a photon is subject to the
 $i$-th
 scattering, $m_p$ is the proton mass, $\sigma_{\rm KN}$ is the
 Klein-Nishina cross section (Rybicki \& Lightman 1979), and the
 integral of $dl$ should be
 done along the photon direction there from the point
 $(x_i,~y_i,~z_i)$ to the boundary of the calculating
 box.  The quantity of $w_0P_0$ represents the transmitted portion
 of photons
 and is recorded to calculate the penetrated spectrum if the path
 of the photon
 does not cross with the equatorial plane, and will no longer
 continue to be
 counted and will be regarded to be absorbed by the disk.  As for
 the remaining
 portion of a photon, its weight becomes $w_1=w_0(1-P_0)$.
  The transmitted portion
 of photons after $i$-th scattering, $w_i P_i$ is recorded to
 calculate the emergent
 spectrum, and the remaining portion, $w_i(1-P_i)$, undergoes the
 $(i+1)$-th scattering.
  This calculation is continued until the weight $w_i$ becomes
 sufficiently small
 ($w_i\ll 1$) or the path of the remaining photon crosses the
 equatorial plane, regarding
 to be absorbed by the disk.  The whole process is simulated by the
 Monte Carlo method.
  Finally, we suppose that the inner boundary of the underlying
 standard disk is $r=3r_{\rm S}$,
 and general relativistic effects like the light bending and
 energy shift are neglected in
 our study.

\section{Results}
\subsection{Spectral Features}
First, we show the emergent spectra from the accretion disk with
 MHD coronal flow with various
 density parameters for the corona in Fig. 1.  In this calculation,
 the mass accretion rate of the underlying disk is set to be
 $\dot{M}_{\rm disk}=10^{-3}M_{\rm Edd}$.  This value is relatively
 lower than that assumed in the standard picture
 ($\dot{M}_{\rm disk}\sim 0.1-0.01M_{\rm Edd}$).  However,
 according to the idea of Haardt \& Maraschi (1991), the liberated
 gravitational energy accompanying the mass accretion $L_G$ should
 be mostly dissipated in
 the corona ($fL_G$, where $f\simeq 1$), while only a small fraction
 of the energy ($(1-f)L_G$) is dissipated in the underlying disk.
  Then the temperature of
 the intrinsic disk radiation would be reduced by a factor of
 $(1-f)^{1/4}$.  In our calculation the mass accretion rate of the
 disk only appears when
 giving the intrinsic disk temperature Eq. (\ref{seed}), and so the
 small mass accretion rate corresponds to the reduced disk temperature.
  Due to
 this fact, the relatively small mass accretion rate adopted in our
 calculation is justified.

\begin{figure*}[ht]
\begin{center}
\FigureFile(130mm,100mm){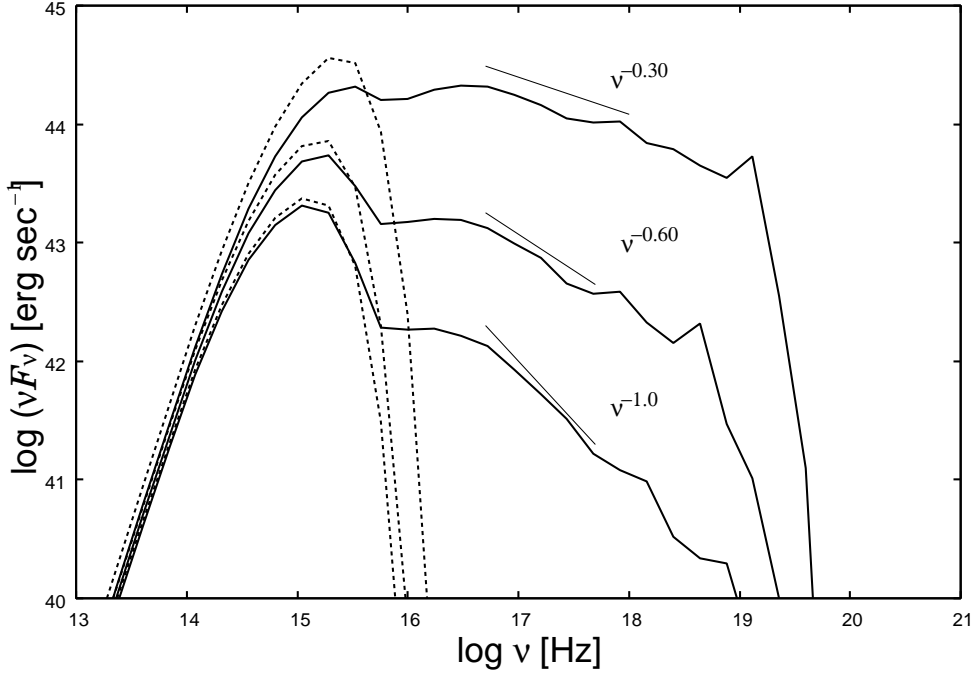}
\end{center}	
\caption{Broadband spectra from a disk at the accretion
 rate of $10^{-3}{\dot M}_{\rm Edd}$ and a MHD coronal flow
 around a black hole
 of $10^8M_{\odot}$ with the density normalization parameters
 $5.1\times 10^{-15}$,
 $1.6\times 10^{-14}$ and
 $5.1\times 10^{-14}~{\rm g}~{\rm cm}^{-3}$ {\it solid lines}.  The spectra of
 the seed photons (including those from the cold disk and the
 reflection component; {\it dashed lines}) are also shown
 for the comparison.  The short solid lines near the spectra show the power-law fit of $\nu F_{\nu}$.}
\label{sedrho}
\end{figure*}

  The adopted density parameters $\rho_0$ for the corona
 are $\rho_0=5.1\times 10^{-15},~1.6\times 10^{-14}$, and $5.1\times
 10^{-14}{\rm g}~{\rm cm}^{-3}$.  These values correspond to
 the number density of $\sim 10^9{\rm cm}^{-3}$.
 Such density is consistent with some corona models (see Liu
 et al. 2002, 2003).  Due to the inverse Compton scattering
 of
 in which the soft thermal photons from the underlying disk in the
 corona, the power-law component with the spectral index of
 $\alpha \sim 1-2$ ($F_{\nu}\propto \nu^{-\alpha}$) appears in the
 higher energy band.  In the calculated spectra, we cannot find a
 bump-like structure 
 clearly in the UV/soft X-ray band which is usually seen in typical
 spectra of AGNs.  As $\rho_0$ increases, the total luminosities
 increases, while the
 power-law index decreases.  The corona with higher density (especially
 with higher scattering optical depth) would irradiate the underlying
 disk with
 higher energy flux because larger number of photons originated from the
 disk would gain the energy and be backscattered.  The the disk is
 heatened by
 the corona and so the energy flux of the seed photon field would
 increase.  This is why the total luminosity rises with the coronal
 density.

\begin{table}
\caption{Coronal properties}
\begin{center}
\begin{tabular}{|cccccc|}
\hline
$\rho_0$ (${\rm g}~{\rm cm}^{-3}$)&region&
$T_{\rm cor}~[{\rm K}]$&$\tau$&$y$&$\alpha$\\ \hline
$5.1\times 10^{-14}$&$0<r<10r_{\rm S}$&
$\sim 5.2\times 10^9$&$\sim 0.2$&$\sim 0.5$&$\sim 1.30$ \\
&$10r_{\rm S}<r<30r_{\rm S}$&$\sim 3.4\times 10^9$&
$\sim 0.9$&$\sim 3$& \\
&$30r_{\rm S}<r$&$\sim 7.6\times 10^8$&$\sim 0.7$&$\sim 0.5$&
 \\ \hline
$1.6\times 10^{-14}$&$0<r<10r_{\rm S}$&
$\sim 5.2\times 10^9$&$\sim 0.05$&$\sim 0.2$&$\sim 1.60$ \\
&$10r_{\rm S}<r<30r_{\rm S}$&$\sim 3.4\times 10^9$&
$\sim 0.25$&$\sim 0.55$& \\
&$30r_{\rm S}<r$&$\sim 7.5\times 10^8$&$\sim 0.2$&
$\sim 0.1$& \\ \hline
$5.1\times 10^{-15}$&$0<r<10r_{\rm S}$&
$\sim 5.2\times 10^9$&$\sim 0.02$&$\sim 0.05$&$\sim 2.00$ \\
&$10r_{\rm S}<r<30r_{\rm S}$&$\sim 3.4\times 10^9$&
$\sim 0.09$&$\sim 0.15$& \\
&$30r_{\rm S}<r$&$\sim 7.6\times 10^8$&$\sim 0.07$&
$\sim 0.03$& \\
\hline
\end{tabular}
\end{center}
\end{table}

In Table 1 we summarize the scattering optical depths $\tau$
 (evaluated by integrating the scattering
 opacity over the z-direction from the equatorial plane), and the
 Compton $y$ parameters of the corona (averaged in each region), and the spectral indices of the
 power law component estimated from the calculation results for 3
 density parameters.  The MHD simulation of coronal flow was
 started from the initial condition of a magnetized torus located
 around at
 $r\sim 40r_S$.  The radius of maximum gas density
 (and of maximum $\tau$) decreases inward with time and stays around
 $r\sim 20r_S$.
  According to the theory of unsaturated inverse Compton scattering,
 the power-law index of the Comptonized emission component depends on
 the plasma
 density and temperature through this equation:
\begin{eqnarray}
\alpha=-\frac{3}{2}+\sqrt{\frac{9}{4}+\frac{4}{y}}. \label{ypara}
\end{eqnarray}
Here $y\equiv (4k_B T/m_e c^2)\tau$ is the Compton $y$-parameter
 (Rybicki \& Lightman 1973), where $T$ and $\tau$ are the temperature
 and
 Thompson optical depth of the corona, respectively.  The spectral
 indices derived from this equation using $y$-parameters in Table 1 do
 not always agree with those estimated from the spectra.  This is
 because $\tau$ and $T_{\rm cor}$ have spatial distributions and we
 cannot evaluate
 $y$-parameter of the corona uniquely.  Even with such a situation,
 however, we can see the tendency that the spectra get flatter with
 higher
 coronal density (and then higher $y$-parameter), which is consistent
 with the theory above.

  In our calculation, $\rho_0$ is determined so as to reproduce the
 observation.
  By tuning the coronal density parameter, we can reproduce the
 luminosity
 of the power-law component which is as intense as that of the thermal
 component originated from the optically thick disk.  Such
 feature is typical in Seyfert galaxies.
  However, the coronal temperature cannot be chosen freely but should
 be determined by imposing the energy balance of the electrons
 between Coulomb collisions and the cooling via inverse Compton
 scattering, as we have done.  Nevertheless the resulting coronal
 temperature
 is substantially reduced from the plasma temperature derived by the
 simulation ($\sim 10^{13}~{\rm K}$) to $\sim 10^9{\rm K}$, which makes
 the high energy cutoff of computed spectra consistent with
 observations.

\subsection{Time variation}
The spectral variation caused by the time variation of MHD coronal flow
 structure is shown in Fig. 2.  In the highest energy range
 ($\gtrsim 10^{18}{\rm Hz}$) the spectra show fluctuations because of
 poor photon statistics.  As for the soft X-ray band
 (with ${\rm log}\nu\simeq 17-18$)
 where the spectra show a smooth power-law shape, the spectral index
 slightly changes with time, and then the X-ray flux fluctuates a little
 (see also Fig. 3).  According to the MHD simulation on which
 our radiative transfer calculations are based, the three-dimensional
 structure of the coronal accretion flow is fluctuating everywhere in
 each timestep.
  On the other hand, the spectral index depends on the distribution of
 $y$-parameter of the corona, as we note in the last subsection.  So we
 can conclude
 that the fluctuations of the spectral indices of the computed spectra
 in Fig. 2 reflect the fluctuation of $y$, which comes from the density
 fluctuations (which is supposed to be due to MRI) in the coronal flow.

\begin{figure*}[ht]
\begin{center}
\FigureFile(130mm,100mm){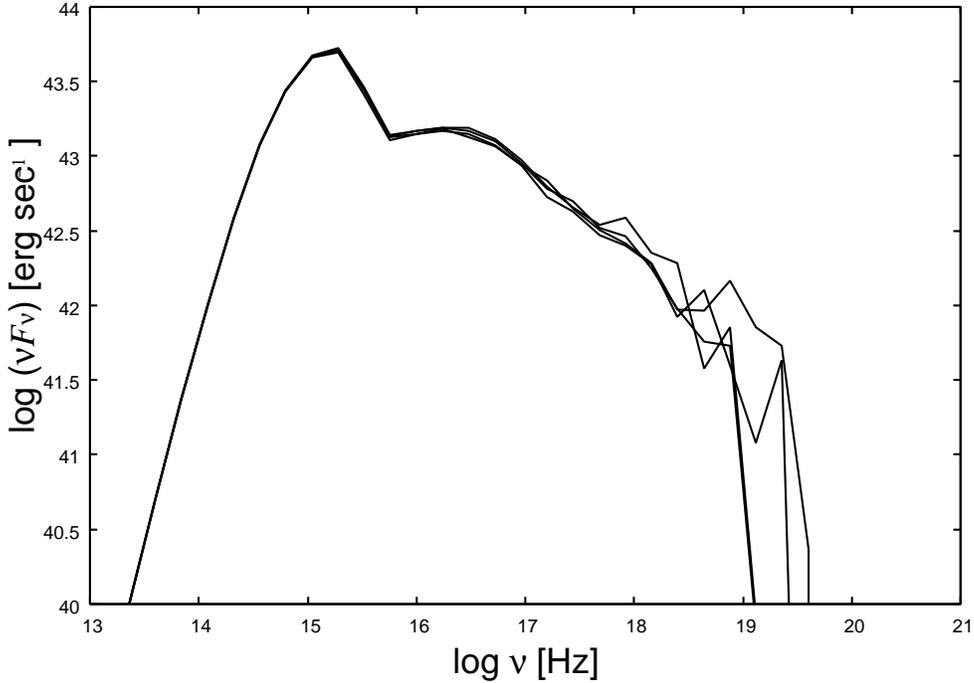}
\end{center}
\caption{Spectral variation of the Comptonized emission
 predicted from the standard disk with a MHD coronal flow
 around a black hole of
 $10^8M_{\odot}$.  Here we set the density parameter as
 $1.6\times 10^{-14}{\rm g}~{\rm cm}^{-3}$.}
\label{sedtime}
\end{figure*}

  Fig. 3. shows the X-ray lightcurve derived from our simulations.
  From this plot we can see that the X-ray luminosity from our
 disk-corona
 system can change by factors of a few tens of percent on timescales of
 the orbital period at the last stable orbit ($r=3r_{\rm S}$),
 i.e. about $10^3(M/10^8) M_{\odot}$ sec.  In the whole calculation
 we do not vary the properties of the soft photon source (i.e. the
 underlying cold disk).  This variation which we obtained
 is due to the density fluctuation (and accompanying temperature
 fluctuation) of the coronal flow.

\begin{figure*}[ht]
\begin{center}
\FigureFile(130mm,100mm){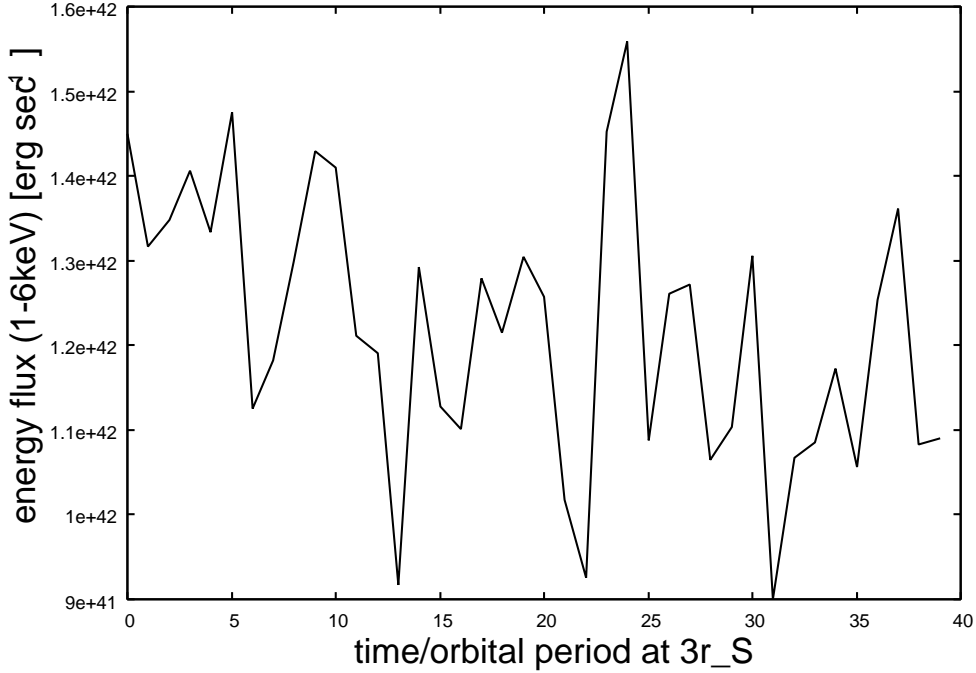}
\end{center}
\caption{Time variation of the X-ray luminosity (1-6keV).
  Here we set the density parameter as
 $1.6\times 10^{-14}{\rm g}~{\rm cm}^{-3}$.}
\label{xlc14}
\end{figure*}

\section{Discussion and Conclusion}
We have calculated the emergent spectra and their time variabilities
 predicted based on the disk-corona model, in which
 a cold standard disk at the equatorial plane is sandwiched by a hot
 coronal flow. As for the structure and
 dynamics of the corona we use the three-dimensional MHD simulation data
 by Kato (2004).  In this section we discuss our results
 especially in the context of AGNs.

We have shown that our black hole disk-corona system can reproduce the
 power-law X-ray emission with the photon index $\alpha\sim 1-2$ by
 adjusting the
 density parameter properly.  The power-law indices and the cutoff
 energy scales of the spectra are roughly
 in agreement with the observed spectra of Seyfert galaxies.  Moreover,
 we find significant variability of the power-law X-ray emission.
  The power-law X-ray emission flux predicted from our model changes
 by a few tens of percent on timescales of the orbital period
 near the last stable orbit, which is about
 $10^3(M/10^8M_{\odot})~{\rm sec}$,
 while the power-law index nor the cutoff energy do not change
 considerably. This variability comes purely from the fluctuation
 of the coronal flow around $r\sim 20r_S$,
 where the scattering optical depth of the coronal flow
 attains its largest value in its structure.  This fluctuation is
 driven by the turbulence as a result of MRI, and its amplitude is large
 enough to explain the observed X-ray variability in Seyfert galaxies
 (e.g. Miniutti et al. 2007).  In UV/soft X-ray band, however, we cannot
 see the bump-like structure which is often found in the spectra of AGNs.
  The same problem occurred in other disk-corona models
 (Shimura et al. 1995; Liu et al. 2003).
  This bump is supposed to be the thermal radiation coming directly from
 the optically thick disk,
 while in those disk-corona models the disk is wholly covered with corona.
  In order to reproduce the
 UV/soft X-ray bump with a model, the coronal structure may need to be
 patchy or locally concentrated.  In the present study we do
 not consider the viewing angle dependence of the spectra.  If we
 observe this accretion flow system face-on, then the scattering
 optical depth along the line of sight would be smaller than in the
 case observed with non-zero viewing angle and the optically thick
 component should be observed more clearly.

Iron K$\alpha$ fluorescent line emissions would occur when the disk is
 irradiated by X-ray from the corona.  The profile of this line emission
 would
 be broadened due to relativistic effects (see Reynolds \& Nowak 2003 for
 a review).  The detailed structure of this line profile is determined by
 the line emissivity distribution on the disk, which depends on the
 spectra of local X-ray irradiation from the corona.

Recently {\it Suzaku}
 has observed the broadened iron line profile from MCG-6-30-15 in detail
 and they implied that the emissivity profile should be as steep as
 $F_{\rm line}(r)\propto r^{-4.4}$ (Miniutti et al. 2007).  This is not
 easy to understand, since standard disks are known to produce
 continuum flux $F_{\rm cont}(r)\propto r^{-3}$.  Some authors proposed
 that this is a manifestation of Kerr black hole effects
 (e.g. Wilms et al. 2001).
  Such a steep emissivity profile can
 be reproduced by thermal Comptonization in the corona, if the
 irradiating corona has the temperature and density profiles which are
 arising inward, as was pointed out by Kawanaka et al. (2005). This is
 because the fraction
 of high energy photons increases inward so that the number of
 irradiating photons which are capable of producing iron fluorescent
 line emissions should increase inward more rapidly than $r^{-3}$.
  As for the MHD corona model which we adopted in this study, however,
 the temperature and density
have flatter profiles (see KMS04), so such a steep line emissivity
 profile is not expected as long as we consider only Compton
 upscattering to be the X-ray
 emission process in this corona.

One of the way of producing such a
 steep emissivity profile so as to produces the iron line profile as
 is observed is to introduce moderate radiative cooling in the
 simulation (Mineshige et al. 2002).
  Radiative cooling can be efficient in dense region which is
 generally located in the inner part of the disk. Then the density
 profile near the
 event horizon of the coronal flow can be steeper and Compton
 up-scattering would be more efficient. As a result, hard X-ray
 photons irradiating the disk would increase and iron fluorescence
 line photons would become
 more efficient. Another possibility is that the transient X-ray
 emission accompanying with magnetic reconnection flares in the
 localized region of the corona would contribute to the iron line
 emissions from the disk,
 which is not taken into account in our present calculation. Such
 small-scale X-ray sources are often assumed when explaining the
 constant reflection component including iron line emissions
 (Miniutti \& Fabian 2004; Malzac et al. 2006; Nayakshin 2007).
  If we analyze the local
 electron heating (or nonthermal acceleration) around magnetic
 reconnection flares and hard X-ray emission processes in the MHD
 coronal flow simulation,
 then we can know the local X-ray irradiation onto the disk, which
 may lead us to understanding the observed broad iron line profiles
 and their time variabilities from the microphysics of X-ray
 emission processes.

We can check the consistency of the assumption that the MHD coronal
 flow which we adopt in the calculation is cooled dominantly by
 advection, by comparing the heating rate ($\sim GMm_p/(2r)\Omega$)
 and the cooling rate ($\sim$Comptonization luminosity per electron).
  According to the result of our calculation, the innermost region
 ($0<r\lesssim 20r_{\rm S}$) is not cooled efficiently by inverse Compton
 scattering.  However, in the outer region ($r\gtrsim 20r_{\rm S}$)
 local heating rate and cooling rate are comparable, which means that
 the assumption of an advection-dominated flow is only marginally justified.
  Generally when Compton cooling becomes efficient the coronal
 temperature will become lower, and then Compton upscattering will be
 inefficient.  This means that the power-law X-ray emission would not
 be generated so much from that region.
  As far as we consider the spectrum and energy flux in X-ray and
 higher energy band, to which the photons from the innermost region
 mainly contribute, this inconsistency would hardly affect the results. 

Finally, we should mention the interaction between the corona and
 the underlying disk.  In the transition zone between the hot corona
 and the
 cold disk where density and temperature abruptly change, the heat
 conduction and
 the mass evaporation would be important as the mass and energy
 exchanging processes (Meyer \& Meyer-Hofmeister 1994;
 Liu et al. 2002).  Evaporation of photospheric material was
 actually shown to be essential in the context
 of solar flares (see e.g. Yokoyama \& Shibata 1994).  By including
 these effects in the simulation we will be able to obtain
 a more realistic model of the coronal structure and dynamics.  To
 what extent the disk-corona structure can extend
 to the inner region is important when we consider the relativistic
 skewing of the iron line emission profile because most of the
 observations of iron line profiles imply that the iron fluorescence
 line photons are
 emitted from around/inside the last stable orbit.  The detailed
 analysis
 of such disk-corona interactions and their observational
 implications are left as a future work. 

\bigskip
We would like to thank Masayuki Umemura, Toshihiro Kawaguchi
 and Ken Ohsuga for helpful comments and discussions.  We are also
 grateful to an anonymous referee for his/her valuable comments, which
 helped us to improve the manuscript in a great deal.
  The numerical calculations were carried out on Altix3700 BX2
 at YITP in Kyoto University.  This work is supported in part by
 Research Fellowship of the Japan Society for the Promotion of
 Science for Young Scientists (N. K.).

\end{document}